\documentclass[twocolumn,showpacs,preprintnumbers,amsmath,amssymb]{revtex4}
\usepackage{epsfig}
\usepackage{graphicx}
\usepackage{dcolumn}
\usepackage{bm}

\begin{document}

\title{Quantum Nonlinear Sigma Model for Arbitrary Spin
Heisenberg Antiferromagnets}
\author{Jianhui Dai$^{a}$ and Wei-Min Zhang$^{a,b,c}$}
\affiliation{$^a$Zhejiang Institute of Modern Physics, Zhejiang
        University, Hangzhou 310027, China\\
 $^b$Department of Physics, National Cheng Kung University,
        Tainan 701, Taiwan \\
 $^c$National Center for Theoretical Science, Taiwan}
\date{19 July, 2005}

\begin{abstract}
In this Letter, we derive a quantum nonlinear sigma model (QNLSM)
for quantum Heisenberg antiferromagnets (QHA) with arbitrary $S$
(spin) values.  A upper limit of the low temperature is naturally
carried out for the reliability of the QNLSM. The $S$ dependence
of the effective coupling constant and the spin wave velocity in
the QNLSM are also obtained explicitly. The resulting spin wave
velocity for 2-dim spin-1/2 QHA highly concurs with the
experimental data of high $T_c$ compound La$_2$CuO$_4$. The
predicted correlation lengths for 2d QHA and spin-gap magnitudes
for 1d QHA also agrees with the accurate numerical results.
\end{abstract}

\pacs{75.10.Jm, 75.40.Gb}

\maketitle

Quantum magnetism in low dimensional strongly correlated systems
is a central issue in modern condensed matter physics. For
example, the parent compounds of the cuprate high $T_c$
superconductors are essentially antiferromagnetic Mott insulators
described by a two-dimensional spin-${1\over 2}$ QHA \cite{mak98}.
Such a low-dimensional QHA system has not been fully solved
quantum mechanically. Only in the {\it large S limit} has Haldane
shown that the lattice QHA can be described by the QNLSM
\cite{haldane}. Since then, the QNLSM has become a good candidate
for the {\it phenomenological} description of the low-dimensional
QHA at low temperatures for various $S$
values\cite{Chak89,HN91,Chub94}.

However, two crucial questions remain for the QNLSM approach: (1)
Why are the predictions of QNLSM obtained in the large-$S$ limit
coincidentally consistent with experimental data of the low energy
QHA with small $S$ values? and (2) What is the upper limit of the
low temperature for the reliability of QNLSM? These two questions,
along with other difficulties for the QNLSM, have been addressed
in some recent literatures\cite{Cuccoli}. But they have not yet
been satisfactorily solved in a simple, consistent approach. In
this Letter, with the topologically invariant spin variable path
integral approach, we resolve these problems by deriving a QNLSM
from the lattice QHA for arbitrary $S$ values.

Let us first briefly recall the large-$S$ approach of the QNLSM.
The partition function of a spin system is usually expressed in
terms of the spin variable path integral as follows (in the unit
$\hbar=k_B=1$):
\begin{eqnarray} \label{cspi}
   Z= Tr e^{-\beta H}=\int {\cal D}[{\bf \Omega}]e^{
        \int_0^\beta d\tau \{i S {\bf A}
        \cdot \dot{{\bf \Omega}}-\langle {\bf \Omega} |
        H |{\bf \Omega} \rangle \}},
\end{eqnarray}
where the first term in the exponent $iS \int_0^\beta d\tau {\bf
A} \cdot \dot{{\bf \Omega}}=\int_0^\beta d\tau\langle {\bf
\Omega}| {d\over d\tau}|{\bf \Omega} \rangle \equiv iS\omega({\bf
\Omega})$ is a topological Berry phase \cite{Wilczek}, and ${\bf
A}$ is a U(1) monopole potential; $|{\bf \Omega}\rangle$ is a spin
coherent state while ${\bf \Omega}$ is a unit vector along which
the spin operator with spin quantum number $S$ is maximally
aligned in $|{\bf \Omega}\rangle$. For the lattice Heisenberg
model (HM), $H=J\sum_{<ij>}{\bf S}_i \cdot {\bf S}_j$ $(J>0)$,
Eq.(\ref{cspi}) becomes
\begin{equation} \label{hmpi}
    Z_H =\int {\cal D}[{\bf \Omega}_i]e^{
        \int_0^\beta d\tau \{iS\sum_i {\bf A}_i \cdot
        \dot{{\bf \Omega}}_i- JS^2 \sum_{<ij>}
        {\bf \Omega}_i \cdot {\bf \Omega}_j\} } .
\end{equation}
By minimizing $H({\bf \Omega})= JS^2\sum_{<ij>}{\bf \Omega}_i\cdot
{\bf \Omega}_j$, one can find the classical ground state (N\'{e}el
state) which spontaneously breaks the SO(3) symmetry. Then by
expanding the action around the ground state, the spin-wave theory
(SWT) of HM, which describes the long wavelength spin
modes\cite{Auerbach}, can be easily derived.

But, according to the Mermin-Wagner's theorem\cite{mwt}, no
symmetry can be spontaneously broken in one- or two-dimensional HM
for a finite $T (> 0)$. To derive an effective long wavelength
action that retains the full spin rotational symmetry, Haldane
considered the large-$S$ limit. In the large-$S$ limit, the path
integrals of Eq.(\ref{hmpi}) are dominated by the semiclassical
equation: $iS{\bf \Omega}\times\dot{{\bf \Omega}} = {\partial
H[{\bf \Omega}]\over
\partial {\bf \Omega}}$. By separating the semiclassical solution
${\bf \Omega}_i$ into a slowly varying N\'{e}el order unit vector
$(-1)^i {\bf n}(x_i)$ plus a slowly varying magnetization density
field perpendicular to ${\bf n}(x_i)$ (Haldane's mapping), and
then taking the continuous limit and integrating out the magnetic
density field, Haldane shows that Eq.(\ref{hmpi}) is reduced to a
QNLSM,
\begin{equation} \label{nlsm}
   Z_H = \int {\cal D}[{\bf n}]~ e^{i2\pi S\Theta[{\bf n}]}
        ~e^{-(\Lambda^{d-1}/2g_0)\int d^{d+1}x~
            \partial_\mu {\bf n} \partial^\mu {\bf n}},
\end{equation}
defined in the $d$+1-dimensional space $(x^1$, $\cdots$,
$x^{d+1})$=$(x^1, \cdots, x^d, c_o\tau)$. Where, $g_o=2\sqrt{d}/S$
is a dimensionless coupling constant, $c_o=2\sqrt{d}JSa$ the spin
wave velocity and $\Lambda=a^{-1}$ the inverse of the lattice
spacing. The imaginary time (temperature) variable $\tau$ ranges
from $0$ to $ \beta =1 /T$. The exponent $\Theta[{\bf n}]$ in
Eq.(\ref{nlsm}) is a topological factor associated with the Berry
phase.

However, as we will elaborate, the derivation of the QNLSM based
on the large-$S$ expansion should be improved from the very
beginning. As a long-standing problem in the construction of
generalized phase space path integrals\cite{Schu}, the
Eq.(\ref{cspi}) is not well defined. The main problem arises from
the assumption, used in deriving Eq.(\ref{cspi}), that $|{\bf
\Omega}(\tau+\delta\tau)\rangle-|{\bf \Omega}(\tau)\rangle$ is of
order $O(\delta\tau)$. Although this assumption has been widely
used in the application of generalized phase space path integrals,
it has never been justified\cite{Schu}. In fact, in effective
field theories, there always exist simultaneous rapidly and slowly
varying paths in the path integral formalism that are associated
with short and long range quantum fluctuations, respectively. The
effective action for slowly varying motions can be properly
obtained by integrating over short range quantum
fluctuations\cite{Wilson}. However, in Eq.(\ref{cspi}), only
slowly varying motions are retained; the short range quantum
fluctuations have been simply ignored.

To overcome the shortcomings involved in the derivation of
Eq.(\ref{cspi}), we begin with the discrete form of the partition
function obtained exactly from the coherent state
representation\cite{wzhang90}:
\begin{eqnarray} \label{dispi}
Z=\lim_{N\rightarrow \infty}\prod_{k=1}^{N}d\mu(
        {\bf \Omega}^k) \exp\{\sum_{k=1}^N \ln
        \langle {\bf \Omega}^k|{\bf \Omega}^{k-1}
        \rangle\nonumber\\-\epsilon {\langle
        {\bf \Omega}^k | H | {\bf \Omega}^{k-1}\rangle
        \over \langle {\bf \Omega}^k|{\bf \Omega}^{k-1}
        \rangle}\},
\end{eqnarray}
where $|{\bf \Omega}^N\rangle=| {\bf \Omega}^0\rangle$ because of
the periodicity of the trace), $\epsilon=\beta/N$ is infinitesimal
as $N\rightarrow \infty$. The slowly varying motion means that
$|{\bf \Omega}_i\rangle - |{\bf \Omega}_{i-1}\rangle$ varies
smoothly in the interval $\epsilon$ such that it can be written as
a time derivative $\epsilon |\dot{{\bf \Omega}}\rangle$. The
rapidly varying motion of $|{\bf \Omega}_i\rangle -|{\bf
\Omega}_{i-1}\rangle$ in the interval $\epsilon$ is related to the
short range quantum fluctuations, we label it as $\delta{\bf
\Omega}$. The assumption of $|{\bf \Omega}^k\rangle- |{\bf
\Omega}^{k-1}\rangle$ being of order $0(\epsilon)$ only keeps the
slowly varying motions $\dot{{\bf \Omega}}$, while the short range
fluctuations $\delta{\bf \Omega}$ is ignored. Thus, the
continuous-time limit of (\ref{dispi}) results in the conventional
spin variable path integral of Eq.(\ref{cspi}).

To include the contribution of the short range fluctuations, one
should expand the nearby coherent state overlap to the second
order terms that either are exclusively slowly varying motions or
include at least one rapidly varying motion. The topologically
invariant terms $\lim_{N\rightarrow \infty} \sum_{k=1}^N \ln
\langle{\bf \Omega}_k|{\bf \Omega}_{k-1} \rangle$ can then be
uniquely expressed by
\begin{eqnarray} \label{geom}
        S\int_0^\beta d\tau \Big\{i {\bf A} \cdot
        \dot{{\bf \Omega}}+ i{\bf \Omega}\cdot
        (\dot{{\bf \Omega}}\times \delta {\bf \Omega})
        - {1\over \tau_\Lambda}\delta {\bf \Omega}\cdot
        \delta {\bf \Omega} \Big\} ,
\end{eqnarray}
where the parameter $\tau_\Lambda \equiv 1/T_\Lambda$ is an
intrinsic shortest time scale (an upper limit of low temperature)
for distinguishing the slowly varying and rapidly varying motions.
We will later discuss this timescale in detail. The second and
third terms in Eq.(\ref{geom}) are usually ignored in the
conventional derivation of phase space path integrals. For the
Hamiltonian term in Eq.(\ref{dispi}), $\lim_{N \rightarrow \infty}
\sum_{k=1}^N \epsilon
    {\langle{\bf \Omega}^k | H | {\bf \Omega}^{k-1}
    \rangle \over \langle {\bf \Omega}^k|
    {\bf \Omega}^{k-1}\rangle}$, since it is already
proportional to $\epsilon$, we only keep the off-diagonal
expansion up to the quadratic order of $\delta {\bf \Omega}$:
\begin{eqnarray}
\int d\tau
    \Big\{H[{\bf \Omega}]
+ {\partial H[{\bf \Omega}]\over \partial
    {\bf \Omega}} \cdot \delta {\bf \Omega} +
    {\partial^2 H[{\bf \Omega}]\over
    \partial {\bf \Omega}_\alpha\partial
    {\bf \Omega}_{\alpha'}} \delta {\bf \Omega}_\alpha
    \delta{\bf \Omega}_{\alpha'} \Big\},
    \label{dyna}
\end{eqnarray}
where $\alpha, \alpha'$ are indices of spin components.
Substituting Eqs.(\ref{geom}-\ref{dyna}) into Eq.(\ref{dispi}),
one gets
\begin{eqnarray}
Z&=&\int {\cal D}[{\bf \Omega}] {\cal D}[\delta
    {\bf \Omega}]~ \exp \int_0^\beta d\tau
    \Big\{iS{\bf A} \cdot \dot{{\bf \Omega}}
    -H[{\bf \Omega}] \nonumber \\
&& ~~~~~~~~~~~ + \Big[ iS{\bf \Omega}\times
    \dot{{\bf \Omega}}-{\partial H[{\bf \Omega}]
    \over \partial {\bf \Omega}}
    \Big] \cdot \delta {\bf \Omega} \nonumber \\
&& ~~~~~~~~ - \Big[{S\over \tau_\Lambda}
    \delta_{\alpha{\alpha'}} + {\partial^2
    H[{\bf \Omega}]\over \partial
    {\bf \Omega}_\alpha \partial
    {\bf \Omega}_{\alpha'}}\Big] \delta
    {\bf \Omega}_\alpha \delta
    {\bf \Omega}_{\alpha'}  \Big\} , \label{cspie}
\end{eqnarray}
which describes both the slowly varying motion $\dot{{\bf
\Omega}}$ and the short range fluctuations $\delta{\bf{\bf
\Omega}}$.

If one takes the semiclassical limit, $iS{\bf
\Omega}\times\dot{{\bf \Omega}} - {\partial H[{\bf \Omega}]\over
\partial{\bf \Omega}}=0$, Eq.(\ref{cspie}) is simply a variation
expansion of the path integral (\ref{cspi}) with respect to the
semiclassical dynamics Haldane used\cite{haldane}, except for the
geometrical term ${S\over \tau_\Lambda}$ which cannot appear in
Haldane's mapping. However, we must emphasize that
Eq.(\ref{cspie}) is derived by carefully treating the off-diagonal
elements of nearby coherent states in Eq.(\ref{dispi}) in terms of
the short range fluctuation $\delta{\bf \Omega}$ and the slowly
varying motions $\dot{{\bf \Omega}}$. Since there is no
semiclassical expansion to begin with, it is not necessary to take
the semiclassical limit by letting the second term vanish.
Instead, one can integrate out the short range fluctuations
$\delta{{\bf \Omega}}$ and obtain a low energy effective action
for the long wavelength spin modes. Since no semiclassical
approximation is made in this procedure, the resulting long
wavelength effective action should be valid for arbitrary $S$
values.

Next, we apply Eq.(\ref{cspie}) to the HM. To specify the
antiferromagnetic ordering, let the slowly varying ${\bf \Omega}_i
=(-1)^i{\bf n}(x_i)$, here the Ne\'{e}l order ${\bf n}(x_i)$ is a
unit vector $|{\bf n}(x_i)|=1$. Then, by taking the space
continuous limit $\sum_i \rightarrow {1\over a^d}\int d^d x$ where
$a$ is the lattice spacing:
\begin{eqnarray}
   && H[{\bf \Omega}] = JS^2 \sum_{<ij>}
    {\bf \Omega}_i \cdot {\bf \Omega}_j \nonumber \\
   &&\rightarrow -dJS^2N + {JS^2\over 2a^{d-2}} \int d^d x
    \sum_{k=1}^d[\partial_k {\bf n}(x)\cdot \partial_k {\bf n}(x)],
    \label{h1} \\
   &&{\partial H[{\bf \Omega}]\over \partial {\bf \Omega}}
    \cdot \delta {\bf \Omega} \rightarrow 0 , \label{csc} \\
   &&{\partial^2 H[{\bf \Omega}]\over \partial {\bf \Omega}_\alpha
    \partial {\bf \Omega}_\beta} \delta {\bf \Omega}_\alpha \delta
    {\bf \Omega}_\beta \rightarrow {2dJS^2 \over a^d} \int d^d x
    \delta {\bf \Omega}(x)\cdot\delta {\bf \Omega}(x).~~~~~~ \label{h3}
\end{eqnarray}
substituting (\ref{h1}-\ref{h3}) into (\ref{cspie}), and
integrating out the short range fluctuation $\delta{{\bf
\Omega}}$, we obtain,
\begin{eqnarray}
    &&Z_H = \int {\cal D}[{\bf n}]~ e^{i2\pi S\Theta[{\bf n}]}
        ~\exp\Big\{- {a^{1-d}\over 2g_s} \int^{c_s\over
        T}_{c_s\over T_{\Lambda}}d(c_s\tau) \nonumber \\
    &&~~~~~~~~~~~~~~~~~~~\times  \int d^d x~\Big[{1\over c_s^2}
        \Big|{\partial{\bf n}\over \partial \tau}\Big|^2
        + |\nabla_x {\bf n}|^2 \Big] \Big\}. \label{qnlsm}
\end{eqnarray}
This is a QNLSM of the low energy QHA for arbitrary $S$ values,
where the coupling constant and spin-wave velocity are given by
\begin{equation}
    g_s={2\over S} \sqrt{d+{T_\Lambda\over 2SJ}}~,~~
    c_s=2JSa\sqrt{d+{T_\Lambda\over 2SJ}}~.  \label{bsc}
\end{equation}
The topological phase factor $2\pi S \Theta[{\bf n}]= 2\pi
S\sum_i(-1)^i{\bf A}(x_i)\cdot\dot{\bf n}(x_i)$ remains the same
as in Haldane's derivation.

In the large-$S$ limit for fixed $T_{\Lambda}$, $g_s\rightarrow
2\sqrt{d}/S, c_s\rightarrow 2\sqrt{d}JSa$. This reproduces the
large-$S$ QNLSM. The difference between (\ref{nlsm}) and
(\ref{qnlsm}) primarily comes from the contribution of the short
range fluctuations, the $1/\tau_\Lambda=T_\Lambda$ term in
(\ref{cspie}) which cannot be included in Haldane's mapping
\cite{haldane}. However, this term plays an important role in the
derivation of a consistent semiclassical dynamics\cite{Klauder}.
Indeed, $T_{\Lambda}$ is an upper limit of the low temperature
scale for the reliability of the QNLSM. Usually one assumes that
there should be no intrinsic cutoff for the imaginary time
variable $\tau$ because quantum fluctuations exist on all time
scales in path integrals\cite{Chak89}. But a low energy effective
theory constructed from path integral is defined by integrating
over high energy quantum fluctuations above certain energy
scale\cite{Wilson}. Without such an intrinsic cutoff, namely, let
$\tau_\Lambda \rightarrow 0$ ($T_\Lambda \rightarrow \infty$),
Eq.(\ref{qnlsm}) reduces to
\begin{equation}
    Z_H \propto \int {\cal D}[{\bf n}]~ e^{i2\pi S\Theta[{\bf n}]}
        ~\exp\Big\{- {\rho_s\over 2T} \int d^d x~
        |\nabla_x {\bf n}|^2 \Big] \Big\}, \label{qnlsm1}
\end{equation}
where $\rho_s=JS^2a^{2-d}$ is the spin stiffness. Except for the
topological phase, this is the classical $d$-dimensional NLSM
rather than a quantum $d+1$-dimensional NLSM that Haldane
obtained\cite{haldane}. This is because in the limit $\tau_\Lambda
\rightarrow 0$, the strong canonical fluctuation in
Eq.(\ref{geom}) smears the dynamical fluctuation of Eq.(\ref{h3})
so that only the classical Hamiltonian Eq.(\ref{h1}) remains.

Now, let us discuss how to consistently determine this timescale.
The lattice spacing $a$ indicates the existence of an intrinsic
momentum cutoff $\Lambda$ in the $d$-dimensional momentum space:
$\Lambda=2\sqrt{\pi}[\Gamma(d/2+1)]^{1/d}/a\equiv L/a$. Using the
energy-momentum relation of the spin wave, $E=c_s k$, one can find
the intrinsic energy cutoff (the inverse of the shortest time
scale $\tau_\Lambda$) $T_\Lambda=c_s\Lambda/2\pi$\cite{Auerbach}.
Combined with (\ref{bsc}), we get
\begin{equation} \label{scale}
    {T_\Lambda \over J}={SL^2\over 4\pi^2}
        \Big(1+\sqrt{1+{16\pi^2 d\over L^2}}~\Big) .
\end{equation}
For $d=2$ and $S=1/2$, we have $L=2\sqrt{\pi}$ and thus
$T_\Lambda/J \simeq 0.97$. This determines quantitatively a low
temperature upper limit for the reliability of QNLSM to the 2-dim
spin-1/2 QHA:  $0 \leq T/J < T_\Lambda/J \simeq 1.0$. Meanwhile,
the spin wave velocity $c_s$ can also be explicitly determined
from Eq.(\ref{bsc}) and Eq.(\ref{scale}). For La$_2$CuO$_4$, which
is a typical 2-dim spin-1/2 QHA with $a=3.79$\AA~and $J \simeq
1500 K$, we obtain (keeping $\hbar$) $\hbar
c_s=2JSa\sqrt{d+{T_\Lambda\over 2SJ}}
        \simeq 0.85~ {\rm eV ~\AA}.$
This is in excellent agreement with the experimental data $\hbar
c_s = 0.85\pm 0.03$ eV \AA \cite{Aeppli}.

Our main results, i.e., Eqs.(\ref{bsc}) and (\ref{scale}), can be
further tested against the known results for the 2d QHA, among
which the quantum Monte Carlo date is almost exact. Up to the
three-loop correction, the QNLSM predicts\cite{HN91} the
asymptotic scaling behavior of the correlation length in the
renormalized classical regime as
\begin{eqnarray}
\xi_{3l}=A\exp(1/t)[1-0.5 t+{\cal O} (t^2)]\label{length},
\end{eqnarray}
where, $A=\frac{e}{8}\frac{\tilde{c}_s}{2\pi\tilde{\rho}_s}$ is a
temperature-independent pre-factor,
$t=\frac{T}{2\pi\tilde{\rho}_s}$ is the dimensionless temperature,
and $\tilde{c}_s$ and $\tilde{\rho}_s$ are the renormalized
spin-wave velocity and spin stiffness which can be consistently
determined by large-$S$ expansion. The predicted formula
$\xi_{3l}$ is extremely sensitive to the spin stiffness and is
consistent with the QMC data\cite{Beard,Kim} at very large
correlation lengths (low temperatures) for $S=1/2$ when the
best-fit value $\tilde{\rho}_s=0.1800$ is used. However, at
moderate correlation lengths, highly accurate QMC data\cite{Beard}
and series expansions\cite{Elstner} indicate a significant
discrepancy which rapidly increases with $S$. The asymptotic
scaling at the three-loop sets in at correlation lengths larger
than $10^5$ for $S=1/2$\cite{Beard} and cosmological lengths for
larger $S$.

\begin{figure}[ht]
\epsfxsize=8cm\epsfysize=4cm \centerline{\epsffile{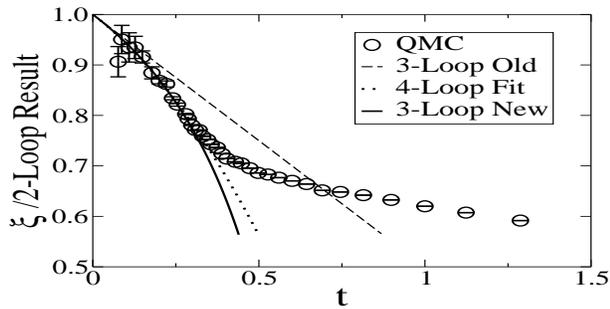}}
\caption[]{Deviations of various results from two-loop asymptotic
scaling as a function of $t$. The QMC result is obtained from
Ref.\cite{Beard}.} \label{scalef}
\end{figure}

Quite strikingly, according to the basic assumption of the
large-$S$ approach, the discrepancy between the theory and
numerics, if it exists, should be significant only for small $S$.
By contrast, in our derivation, the asymptotic scaling behavior
Eq.(\ref{length}) holds for arbitrary $S$ at low temperatures,
provided the effects of the intrinsic scale $T_{\Lambda}$ are
correctly taken into account. Note that the temperature dependence
in Eq.(\ref{length}) comes from a simple assumption that at
nonzero temperature $T$ the correlation length is much larger than
the finite extent $c_s/T$ along the Euclidean time direction in
the renormalized classical regime\cite{Chak89,HN91}. In our case,
the time extent is replaced by $c_s(1/T-1/T_{\Lambda})$, making
the assumption more reasonable. Therefore, we only need to replace
$T$ by the re-scaled temperature $\tilde {T}=\frac{T
T_{\Lambda}}{T_{\Lambda}-T}$ in Eq.(\ref{length}). The validity of
this simple re-scaling requires that $\tilde {T} <
2\pi\tilde{\rho}_s$, or $T<0.5J$ for $S=1/2$. Interestingly, the
re-scaling does not change the two-loop asymptotic scaling
behavior. On the other hand, the shifts in $c_s$ and $g_{s}$ only
modify the pre-factor $A$ but keep $\rho_s$ unchanged. Figure
\ref{scalef} shows the deviations of various results from two-loop
asymptotic scaling $\exp(1/t)$ as a function of $t$. The
three-loop results of Eq.(\ref{length}) in terms of $t$ and
$\tilde{t}=\frac{\tilde {T}}{2\pi\tilde {\rho}_s}$ are plotted as
the $3$-loop old (dashed) line and the $3$-loop new (solid)line,
respectively. For our three-loop result, the scaling regime begins
at roughly $\xi\approx 10^2-10^3$, when $T\approx 0.3J$. As a
comparison, a suggested four-loop (dotted) line\cite{Beard} is
plotted by adding the correction ${\cal O} (t^2)$ in
Eq.(\ref{length}) before re-scaling. The coefficient of this term
is $-0.75$ by fitting the highly accurate QMC data which is
apparently too large to be obtained within reasonable four-loop
corrections\cite{Beard}.

Our results can also be tested against the known results for the
1d QHA.  Note that in 1d, the topological term plays a crucial
role, leading to the quantum critical and disordered phases for
half integer and integer spins,
respectively\cite{haldane,Affleck}.  In the disordered phase,
$T_{\Lambda}$ can be determined by using the relation $E^2=c_s^2
k^2+\Delta_s^2$. This only leads to a correction ${\cal
O}((\frac{\Delta_s}{2SJ})^2)$ to Eq.(\ref{scale}). For
$S=1/2,1,3/2$ and $2$, the accurate numerical results for the
values of $c_s/(aJ)$ obtained by the Density Matrix
Renormalization Group(DMRG)\cite{Hallberg,Lou,Qin} are $1.57,
2.49, 3.87$ and $4.65$ respectively, which show a systematic
deviation from $2S$. While, by Eq.(\ref{bsc}), they are
$1.28,2.55,3.84$ and $5.00$ respectively, providing much better
predictions. The 1d QNLSM also predicts gap\cite{Affleck} in the
disordered phase as $\Delta_s=Bc_s\exp (-2\pi/g_s)$, where $B$ is
a $S$-independent fitting parameter of order of $1$. The DMRG
results are $\Delta_1=0.411J$ for $S=1$\cite{White};
$\Delta_2=0.085J$ for $S=2$\cite{Scholl,Qin}. By using the
best-fitting parameter, $B\approx 2$, we find that our derivation
gives $\Delta_1=0.438J$ and $\Delta_2=0.076J$, respectively, while
the large-$S$ approach gives $\Delta_1=0.172J$ and
$\Delta_2=0.015$, respectively. Therefore, the present 1d QNLSM
provides a better approximation for the spin-gap magnitudes in the
1d integer QHA.

In conclusion, by using a topologically invariant spin variable
path integral approach, we resolve the two crucial problems in the
QNLSM description of QHA as mentioned in the beginning of this
Letter. The basic parameters in the QNLSM are unambiguously
defined for arbitrary spin values. The primary tests discussed
above show that the quantum fluctuations in the QHA, which are
usually underestimated in the large-$S$ approach, are now more
properly described. It should be emphasized that the construction
of a low energy effective field theory from the extended phase
space path integrals developed in this Letter is a general
approach, in which the shortest time scale plays a crucial role
for self-consistency. This approach can be applied to other
generalized phase space path integrals\cite{wzhang90} for the
study of low energy physics in various strongly correlated
systems.

J.D. acknowledges the hospitality of the Abdus Salam International
Centre for Theoretical Physics, Trieste, Italy, where this work
was completed. He would also like to thank Xiaoxuan Huang and
Shaojin Qin for helpful discussions. W.M.Z would like to thank
Heping Ying for the hospitality during his visiting of Zhejiang
Institute of Modern Physics. This work is supported in part by the
NSF of China, the NSF of Zhejiang Province(J.D) and by the NSC
grant No. 93-2112-M-006-019 (W.M.Z).

\end{document}